\documentclass[reprint, aps, prl, amsmath, amssymb, floatfix, superscriptaddress, longbibliography]{revtex4-2}

\usepackage{graphicx}
\usepackage{subcaption}
\usepackage{ragged2e}  

\usepackage{dcolumn}
\usepackage{bm}
\usepackage[utf8]{inputenc}

\usepackage[dvipsnames]{xcolor}
\definecolor{TaskRed}{RGB}{145,0,0}
\definecolor{myblue}{rgb}{0,0,1}
\usepackage[
  breaklinks=true,
  colorlinks=true,
  linkcolor=myblue,
  urlcolor=myblue,
  citecolor=myblue
]{hyperref}

\begin{document}

\title{Surface-bulk magnetic dichotomy in intercalated transition metal dichalcogenides}
\author{Ke Liao}
\email{ke.liao@yale.edu}
\affiliation{Department of Chemistry, Yale University, New Haven, Connecticut 06520,
USA}
\author{Wenxin Li}
\affiliation{Department of Applied Physics, Yale University, New Haven, Connecticut 06511, USA}
\author{Kelly Dai}
\affiliation{Department of Physics, Yale University, New Haven, Connecticut 06520, USA}
\author{Ariel Mazor}
\affiliation{Department of Physics, Santa Clara University, Santa Clara, California, 95053, USA}
\author{Shan Wu}
\affiliation{Department of Physics, Santa Clara University, Santa Clara, California, 95053, USA}
\author{Yu He}
\email{yu.he@yale.edu}
\affiliation{Department of Applied Physics, Yale University, New Haven, Connecticut 06511, USA}
\author{Tianyu Zhu}
\email{tianyu.zhu@yale.edu}
\affiliation{Department of Chemistry, Yale University, New Haven, Connecticut 06520,
USA}

\date{\today}
 
\begin{abstract}
Magnetically intercalated transition metal dichalcogenides are layered materials with highly tunable magnetic orders and unconventional transport properties. Here we report a surface-bulk magnetic dichotomy in \(\mathrm{M}_{1/3}\mathrm{T}\mathrm{X}_{2}\), in which the surface and bulk favor distinct magnetic orders. In \(\mathrm{Fe}_{1/3}\mathrm{Nb}\mathrm{S}_{2}\), angle-resolved photoemission spectroscopy (ARPES) results are reproduced by an A-type antiferromagnetic electronic structure, while neutron scattering and bulk DFT+\(U\) calculations identify stripe order as the bulk ground state. DFT+\(U\) slab calculations resolve this contradiction by showing that relaxation of an Fe-terminated surface reverses the magnetic preference from stripe to A-type order. The same dichotomy recovers the \(K\)-centered Fermi-surface pockets that are observed in \(\mathrm{Co}_{1/3}\mathrm{Nb}\mathrm{S}_{2}\) and \(\mathrm{Co}_{1/3}\mathrm{Ta}\mathrm{S}_{2}\) but absent from previous bulk calculations. Notably, the surface-stabilized magnetic phase of \(\mathrm{Fe}_{1/3}\mathrm{Nb}\mathrm{S}_{2}\) carries nonrelativistic spin splitting and can exhibit altermagnetic character. These results caution against the general association of ARPES signals with bulk magnetic states in these systems, and suggest surface-bulk magnetic dichotomy as a promising route to realize magnetic phases, including altermagnetism, that are not energetically favored in the bulk.
\end{abstract}

\maketitle

\begin{figure*}[htb!]
    \centering
    \includegraphics[width=0.99\textwidth]{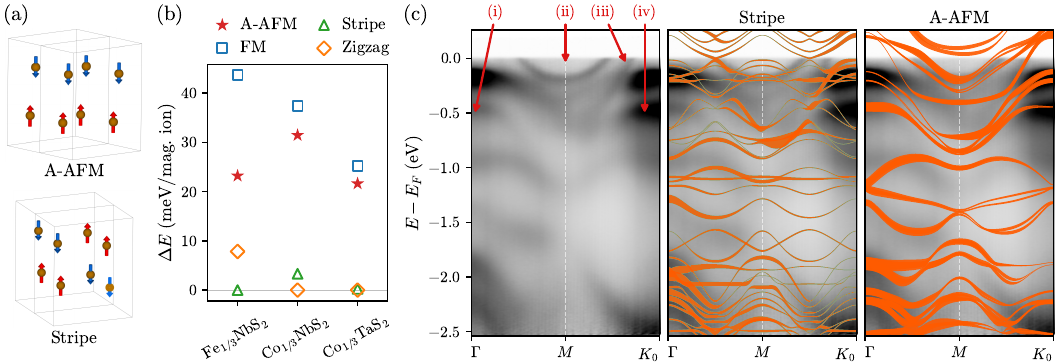}
    \caption{\justifying Magnetic competition and ARPES--DFT comparison in
    \(\mathrm{M}_{1/3}\mathrm{T}\mathrm{X}_{2}\).
    (a) A-AFM and stripe orders on the intercalant sublattice.
    (b) Bulk PBE ($U=0$) energies of collinear orders relative to the lowest order per compound, calculated on FM-optimized structures.
    (c) ARPES spectrum of \(\mathrm{Fe}_{0.32}\mathrm{Nb}\mathrm{S}_{2}\) (left; Ref.~\cite{li2025}), overlaid with bulk DFT+\(U\) (\(U=0.3\)~eV) spin-summed unfolded spectral weight of stoichiometric \(\mathrm{Fe}_{1/3}\mathrm{Nb}\mathrm{S}_{2}\) for stripe (middle) and A-AFM (right) orders at \(k_z=\tfrac{1}{8}\tfrac{2\pi}{c}\), calculated on experimental structure~\cite{wu2022}. Arrows (i)–(iv) mark discrepancies between the stripe calculation and experiment. \(M\) and \(K_{0}\) denote \(\sqrt{3}\times\sqrt{3}\) supercell and primitive zone points.}
    \label{fig:1}
\end{figure*}

Magnetic intercalation transforms metallic transition metal dichalcogenides
(TMDs) into a highly tunable family of layered magnets~\cite{Parkin1980Magnetism}. At commensurate
concentrations, \(3d\) ions form ordered superlattices in the van der Waals (vdW)
gaps of 2H Nb- and Ta-based chalcogenide hosts, coupling local moments to itinerant
quasi-two-dimensional bands~\cite{WilsonYoffe1969AdvPhys,VanLaar1971MagneticStructures,Xie2022JACS}.
Depending on the intercalant species, host, and concentration, these compounds
realize ferromagnetic (FM)~\cite{Parkin1980Magnetism} and helical phases~\cite{Togawa2012CSL} as well as several antiferromagnetic (AFM)
orders, including stripe, zigzag, and A-type (A-AFM) [Fig.~\ref{fig:1}(a)]. These magnetic states are often accompanied by unconventional
transport responses~\cite{lu2020a,an2023,sirica2016,Sirica2020Cr13NbS2,
xie2023,park2023,hatanaka2023,Aoki2019Magnetochiral,ghimire2018}. In a given
\(\mathrm{M}_{1/3}\mathrm{T}\mathrm{X}_{2}\) system, where \(\mathrm{M}\) denotes
the \(3d\) intercalant, \(\mathrm{T}=\mathrm{Nb}\) or \(\mathrm{Ta}\), and
\(\mathrm{X}=\mathrm{S}\) or \(\mathrm{Se}\), different magnetic orders compete closely as the ground state
[Fig.~\ref{fig:1}(b)]. Across this family, the lowest-energy magnetic state
changes both with elemental composition and with small deviations from stoichiometry~\cite{Mangelsen2021CoComposition,wu2022,Park2024CoTaS2Composition,Hawkhead2023BandFilling}. 

Despite the nominal notion of these compounds being intercalated derivatives of vdW hosts, because the natural cleavage plane lies between the intercalant and host
layers, the resulting surface has a substantially modified coordination and
electronic environment from the bulk~\cite{WeberSpaldin2023Surface,Edwards2024ChemMater}. These changes are particularly relevant when comparing
angle-resolved photoemission spectroscopy (ARPES), which typically probes
the near-surface electronic structure~\cite{Sobota2021ARPES}, with neutron diffraction, which
determines magnetic order in the bulk. Previous work has established
substantial electronic reconstruction at several cleaved intercalated-TMD
surfaces due to intercalant clustering~\cite{Edwards2024ChemMater,edwards2023}.
Micro-ARPES and slab calculations on \(\mathrm{Co_{1/4}Nb}\mathrm{Se}_{2}\)
resolve coexisting Co- and Se-terminated domains with distinct surface
electronic structures~\cite{Forcella2026NanoLett}. Surface magnetism has long been known to depart from the bulk. Surfaces can order above the bulk transition temperature~\cite{Vescovo1993,Barbier2004,Guo2024,Pei2024} or host a magnetic phase absent from the bulk~\cite{Matzdorf2000,Pressacco2016,Niu2020,Freeland2005}. Whether the cleaved surfaces of intercalated TMDs can favor a
magnetic order distinct from the bulk remains an open question. A change in surface magnetic order may also alter
the symmetry of the electronic structure~\cite{lange2026}. Among all the competing magnetic orders in 1/3-intercalated
2H TMDs, A-AFM (ferromagnetic intercalant layers stacked
antiferromagnetically along the \(c\) axis) has \(g\)-wave altermagnetic
symmetry~\cite{day-roberts2026,Fender2025JACS,Smejkal2022PRX031042,Smejkal2022PRX040501}. Its stabilization near the surface could therefore allow distinct 2D magnetic phases, including those bearing altermagnetic character, in compounds whose bulk favors a different order.

\(\mathrm{Fe}_{1/3}\mathrm{Nb}\mathrm{S}_{2}\) provides a  test of
this possibility. Its AFM phase hosts current-induced resistive switching~\cite{Nair2019Fe13NbS2Switch,Maniv2021Switching,Haley2023Nonlocal}, three-state electronic nematicity~\cite{Little2020Fe13NbS2}, and magnetoelastic coupling~\cite{kar2025,Wu2023ChargeOrder}. Around the
critical concentration \(x=1/3\), neutron scattering identifies stripe and
zigzag bulk orders on the under- and over-intercalated sides,
respectively~\cite{wu2022}. Bulk calculations likewise place
these orders in close competition, with their relative stability sensitive to correlation strength~\cite{weber2021}. Recent
ARPES measurements traced the bulk stripe-to-zigzag transition to
correlation-driven decoherence of narrow Fe \(3d_{z^2}\) states
hybridized with itinerant Nb bands~\cite{li2025}. Yet the band-by-band agreement between stripe-order magnetic DFT calculations and the measured ARPES spectrum on underintercalated
\(\mathrm{Fe}_{0.32}\mathrm{Nb}\mathrm{S}_{2}\) remains incomplete [Fig.~\ref{fig:1}(c)]. Similar discrepancies occur in cobalt-intercalated TMDs, where calculations based on collinear bulk-order descriptions do not reproduce the prominent \(K\)-centered Fermi-surface pockets observed by ARPES, while noncollinear ordering and correlation-based interpretations have also been proposed~\cite{popcevic2022,Yang2022PRBL121107,park2023,Gu2025NatCommun,Takagi2023TopologicalHall,Dong2024TripleQ,Luo2026CoTaS2VHS}. 

Here we report a possible surface-bulk magnetic dichotomy in which the surface and bulk favor different magnetic orders.
In \(\mathrm{Fe}_{1/3}\mathrm{Nb}\mathrm{S}_{2}\), A-AFM DFT+\(U\)
bands capture the key features of the ARPES spectrum, even though neutron scattering and bulk energetics identify stripe or zigzag order as the bulk ground state. Slab calculations show that an
idealized Fe-terminated surface reverses the energetic preference and stabilizes
A-AFM order. The same A-AFM description of
\(\mathrm{Co}_{1/3}\mathrm{Nb}\mathrm{S}_{2}\) and
\(\mathrm{Co}_{1/3}\mathrm{Ta}\mathrm{S}_{2}\) recovers the \(K\)-centered
Fermi-surface pockets previously missing in bulk magnetic ground state calculations. In \(\mathrm{Fe}_{1/3}\mathrm{Nb}\mathrm{S}_{2}\), the surface-stabilized A-type phase also carries nonrelativistic, sign-changing spin splitting characteristic of altermagnetism.

We first compare the ARPES spectrum of underintercalated
\(\mathrm{Fe}_{0.32}\mathrm{Nb}\mathrm{S}_{2}\) in
Fig.~\ref{fig:1}(c)~\cite{li2025} with band structures of stoichiometric \(\mathrm{Fe}_{1/3}\mathrm{Nb}\mathrm{S}_{2}\) calculated for the stripe and A-AFM magnetic states. For direct comparison with ARPES, the calculated
bands are unfolded onto the primitive \(\mathrm{NbS}_2\) Brillouin zone,
with the corresponding unfolded spectral weights shown. As illustrated in Fig.~\ref{fig:1}(a), A-AFM preserves ferromagnetic order within each intercalant layer and alternates only between adjacent layers, whereas stripe order breaks the in-plane translational symmetry. Unless otherwise stated, all band structures are calculated in VASP~\cite{Kresse1996VASP} with projector-augmented-wave potentials~\cite{Bloechl1994PAW,Kresse1999PAW}, using DFT+\(U\)~\cite{Dudarev1998PlusU} with the PBE exchange-correlation functional~\cite{perdew1996generalized} and \(U=0.3\)~eV applied to the intercalant \(3d\) orbitals~\cite{haley2020half}. For
\(\mathrm{Fe}_{1/3}\mathrm{Nb}\mathrm{S}_{2}\), this choice has been shown to reproduce the experimental magnetocrystalline
anisotropy~\cite{haley2020half,weber2021}.
The A-AFM and stripe calculations in Fig.~\ref{fig:1}(c) use the same experimentally refined structure and DFT+$U$ parameters. Because the detailed band energies depend on \(U\), we assess the agreement qualitatively, focusing on the presence or absence of distinct spectral features. As shown in Fig.~\ref{fig:1}(c), the stripe calculation disagrees with experiment in at least four ways: it (i) misses the band near \(E-E_F\approx-0.5\)~eV at \(\Gamma\), (ii) adds extra $M$-centered hole-like bands, (iii) misses a Fermi-level crossing near
\(K_0\), and (iv) misses the flat bands near
\(E-E_F\approx-0.5\)~eV at \(K_0\). In contrast, the A-AFM calculation reproduces the observed presence or absence of these features and captures the overall low-energy dispersion much more consistently. The weak splitting visible in the calculated A-AFM bands along this cut arises from the slight symmetry lowering of the experimentally refined bulk structure.
The surface-orbital-projected bands of the relaxed A-AFM slab along the same in-plane cut are shown in Fig.~S6 of the Supplemental Material (SM)~\cite{suppmat}.

\begin{figure*}[htb!]
    \centering
    \includegraphics[width=0.99\textwidth]{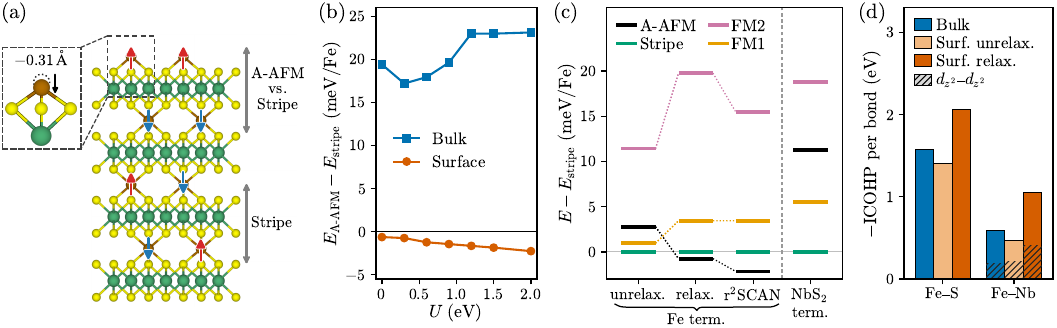}
    \caption{\justifying Surface-bulk magnetic energetics and bonding in \(\mathrm{Fe}_{1/3}\mathrm{Nb}\mathrm{S}_{2}\).
    (a) Fe-terminated slab (green: Nb; yellow: S; brown: Fe). The lower two Fe layers are fixed at bulk positions in stripe order; the upper two are relaxed separately for each magnetic configuration. Inset: the surface Fe moves \(0.31\)~\AA\ inward. (b) A-AFM--stripe energy difference versus \(U\) for the bulk experimental structure~\cite{wu2022} and the Fe-terminated slab (geometries relaxed at \(U = 0.3\)~eV for each magnetic phase). (c) Energies of four surface orders relative to stripe, from PBE+\(U\) (\(U=0.3\)~eV) for the unrelaxed and relaxed Fe-terminated slabs and the relaxed NbS\(_2\)-terminated slab, and from r$^2$SCAN (no \(U\)) on the PBE+$U$--relaxed Fe-terminated geometries. 
    FM1: top Fe layer ferromagnetic, second Fe layer stripe; FM2: top two Fe layers ferromagnetic and parallel.
    (d) \(-\)ICOHP per Fe--S and Fe--Nb bond in the bulk and at the unrelaxed and relaxed Fe-terminated surfaces in the A-AFM order.}
    \label{fig:2}
\end{figure*}

Despite this agreement with ARPES, A-AFM is not favored in the bulk.
Neutron diffraction establishes stripe order for the underintercalated
composition, including $x=0.32$, and zigzag order for
overintercalated compositions,
with both phases competing near stoichiometry~\cite{wu2022}.
Fig.~\ref{fig:1}(b) shows that the energy hierarchy of different magnetic configurations varies across
\(\mathrm{M}_{1/3}\mathrm{T}\mathrm{X}_{2}\), and A-AFM is not generically the lowest-energy state. For stoichiometric
\(\mathrm{Fe}_{1/3}\mathrm{Nb}\mathrm{S}_{2}\), our PBE calculations place stripe below A-AFM by
\(23.2\)~meV/Fe at \(U=0\).
Thus, in Fig.~\ref{fig:1}, surface-sensitive ARPES reports an A-AFM-like electronic structure, whereas bulk neutron diffraction and bulk energetics are more consistent with a stripe AFM order.


This contrast necessitates a careful examination of whether the A-AFM-like state is specific to the surface and, in particular, to its termination.
The features highlighted in Fig.~\ref{fig:1}(c) can be attributed to Fe-terminated surface regions based on low-energy electron diffraction (LEED), spatially resolved x-ray photoelectron spectroscopy (XPS), spatially resolved ARPES measurements, and comparisons to the calculated electronic structures of Fe- and NbS$_2$-terminated surfaces (Fig.~S1 and the accompanying discussion in the SM~\cite{suppmat}). Motivated by this assignment, we use the four-layer, 80-atom slab shown in Fig.~\ref{fig:2}(a) to test the surface-specific energy hierarchy of the leading bulk magnetic configurations.
In the slab model, the lower two
layers are fixed in the bulk stripe order and at the bulk atomic positions,
while the upper two layers are relaxed independently from stripe and A-AFM
initial magnetic conditions. Because both slabs share the same fixed stripe interior, their energy difference measures the relative stability of the near-surface orders. 

In Fig.~\ref{fig:2}(b), we evaluate single-point energies at each $U$ using the experimental bulk structure and the slab geometries relaxed by PBE+$U$ ($U=0.3$~eV) for each magnetic phase. The magnetic preference
\(\Delta E=E_{\mathrm{A\text{-}AFM}}-E_{\mathrm{stripe}}\) remains positive in the bulk (\(+17.2\) to \(+23.1\)~meV/Fe) but becomes negative at the surface (\(-0.6\) to \(-2.3\)~meV/Fe) across \(U=0\)--\(2\)~eV. The Fe-terminated surface therefore reverses the bulk magnetic preference
and stabilizes the ARPES-compatible A-AFM phase, while the interior remains stripe ordered.

Figure~\ref{fig:2}(c) and Table~S6 further compare four magnetic orders of the top two Fe layers. Surface relaxation changes \(\Delta E\) from \(+2.8\)~meV/Fe on the unrelaxed, bulk-terminated geometry to \(-0.7\)~meV/Fe on the separately relaxed A-AFM and stripe geometries. The relaxed configurations with only the top Fe layer
ferromagnetic (FM1) or both upper Fe layers aligned ferromagnetically (FM2) lie about \(4.2\) and \(20.5\)~meV/Fe above A-AFM,
confirming that A-AFM is the lowest-energy state among the four tested orders. Single-point r$^2$SCAN calculations without \(U\) on the PBE+\(U\)-relaxed A-AFM and stripe geometries also favor A-AFM by \(2.2\)~meV/Fe. With the NbS\(_2\) termination, PBE+\(U\) (\(U=0.3\)~eV) instead favors stripe over A-AFM by \(11.3\)~meV/Fe. The A-AFM preference is therefore specific to the Fe termination.

Orbital-resolved ARPES identifies narrow, moment-carrying Fe
\(3d_{z^2}\) states that strongly hybridize with itinerant Nb
\(4d_{z^2}\) bands, providing the axial pathway for interlayer Fe-Fe exchange and the bulk stripe-zigzag transition~\cite{li2025}. At the Fe-terminated surface, the outermost Fe atoms lose their NbS\(_2\)
neighbor on the vacuum side and relax inward by \(0.31\)~\AA,
shortening the Fe--S and Fe--Nb distances from \(2.43\) to
\(2.24\)~\AA{} and from \(3.05\) to \(2.73\)~\AA, respectively
[Fig.~\ref{fig:2}(a)]. Integrated crystal orbital Hamilton population (ICOHP) analysis~\cite{Dronskowski1993COHP}, in which a larger \(-\mathrm{ICOHP}\) value indicates stronger bonding, shows increased \(-\mathrm{ICOHP}\) per bond for both
Fe--S and Fe--Nb after relaxation [Fig.~\ref{fig:2}(d) and Table~S7 in the SM~\cite{suppmat}], indicating stronger bonding to the remaining host layer. The increased Fe \(3d_{z^2}\)--Nb \(4d_{z^2}\) contribution further indicates a change
in the interlayer hybridization pathway. A short-range pairwise Heisenberg model fitted to bulk DFT+\(U\) energies cannot reliably rank
the closest competing orders (Fig.~S5 and Table~S9 in the SM~\cite{suppmat}), suggesting an important role for
long-range or non-pairwise exchange mediated by itinerant Nb states.
Such exchange is sensitive to Fe--Nb hybridization and could therefore account for the reversal
upon surface relaxation.

\begin{figure}[htbp]
    \centering
    \includegraphics[width=\columnwidth]
    {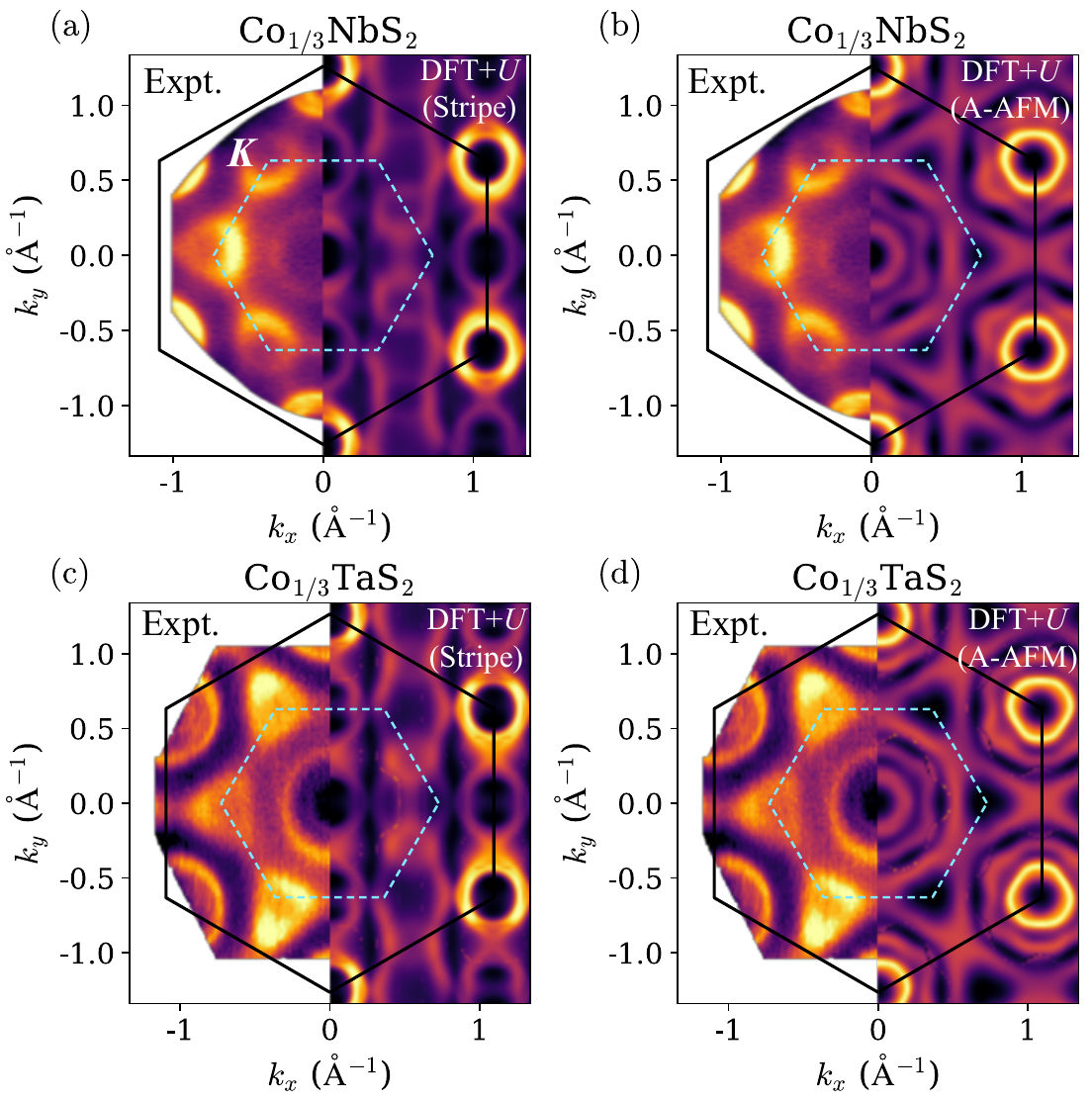}
    \caption{\justifying A-AFM recovery of the \(K\)-centered pockets in
    cobalt-intercalated TMDs.
    (a,b) Fermi surfaces of \(\mathrm{Co}_{1/3}\mathrm{NbS}_{2}\).
(c,d) Fermi surfaces of \(\mathrm{Co}_{1/3}\mathrm{TaS}_{2}\). In each panel, the left half shows the experimental ARPES
intensity and the right half shows bulk DFT+\(U\) (\(U=0.3\)~eV)
unfolded spectral weight for stripe [(a,c)] or
A-AFM order [(b,d)]. The experimental ARPES spectra are adapted from Refs.~\cite{popcevic2022,sas2026}. The solid black and dashed cyan hexagons
denote the primitive and \(\sqrt{3}\times\sqrt{3}\) supercell Brillouin
zones, respectively.}
    \label{fig:co}
\end{figure}

We next ask whether the same surface-bulk dichotomy extends to cobalt
intercalates, where a similar ARPES-DFT mismatch appears. Previous ARPES studies of
\(\mathrm{Co}_{1/3}\mathrm{Nb}\mathrm{S}_{2}\) and
\(\mathrm{Co}_{1/3}\mathrm{Ta}\mathrm{S}_{2}\) show shallow Co-derived states near the \(K\) points of the
\(\sqrt{3}\times\sqrt{3}\) Brillouin zone, forming triangular or
dewdrop-like Fermi-surface pockets
~\cite{popcevic2022,noh2025,sas2026}. DFT and DFT+\(U\)
calculations in the magnetic reference states adopted in those works reproduce
much of the host-derived dispersion but miss these pockets. In contrast, our DFT+$U$ ($U=0.3$ eV)
calculations of \(c\)-axis A-AFM phase recover the \(K\)-centered pockets in both compounds, while the stripe order does not, as shown in Fig.~\ref{fig:co}. We use stripe order as a collinear reference here, as the bulk order of both compounds remains debated among collinear, noncoplanar multi-\(q\), double-\(Q\),
and helical states~\cite{ghimire2018,Tanaka2022,Takagi2023TopologicalHall,Dong2024TripleQ,Zager2026DoubleQ,Park2024CoTaS2Composition,Cho2026AlternatingChirality}. Although the measured pockets are
broader and more anisotropic, their existence and momentum-space location are well reproduced in our A-AFM calculations. 

To determine whether this ARPES-compatible A-AFM state is also favored near the surface, we repeated the slab calculation for
\(\mathrm{Co}_{1/3}\mathrm{Nb}\mathrm{S}_{2}\). Using DFT+\(U\) with
\(U=0.3\)~eV, both stripe and A-AFM surface states remain stable upon
relaxation, with A-AFM lower in energy by \(1.7\)~meV/Co. Thus, as in
\(\mathrm{Fe}_{1/3}\mathrm{Nb}\mathrm{S}_{2}\), the Co-terminated surface reverses the
relative stability of the two collinear states and favors the
A-AFM phase near the surface in the ground state. 

Temperature-dependent ARPES
finds no resolved change in the \(K\)-centered pockets across \(T_N\) in either
compound~\cite{popcevic2022,sas2026}, which
is compatible with a surface order that persists above the bulk transition. Correlation-based approaches can also produce low-energy Co spectral weight near \(K\). A slave-boson-inspired treatment shifts the Co-derived
states toward \(E_F\) by renormalizing the Co-host hybridization
~\cite{popcevic2022}, while paramagnetic
DFT+DMFT produces a weak \(K\)-centered pocket~\cite{noh2025,park2024}.
Cluster perturbation theory (CPT) reproduces the triangular
pocket in \(\mathrm{Co}_{1/3}\mathrm{TaS}_{2}\)
~\cite{sas2026}, which Sas
\textit{et al.} attribute to bulk correlation requiring long-range Co order. The CPT calculation, however, starts from a
primitive-cell DFT+\(U\) Wannier Hamiltonian with oppositely polarized Co
sites, consistent with an A-type-like interlayer magnetic moment arrangement,
although the magnetic state is not explicitly identified~\cite{sas2026}.

\begin{figure}[htb!]
    \centering
    \includegraphics[width=\columnwidth]{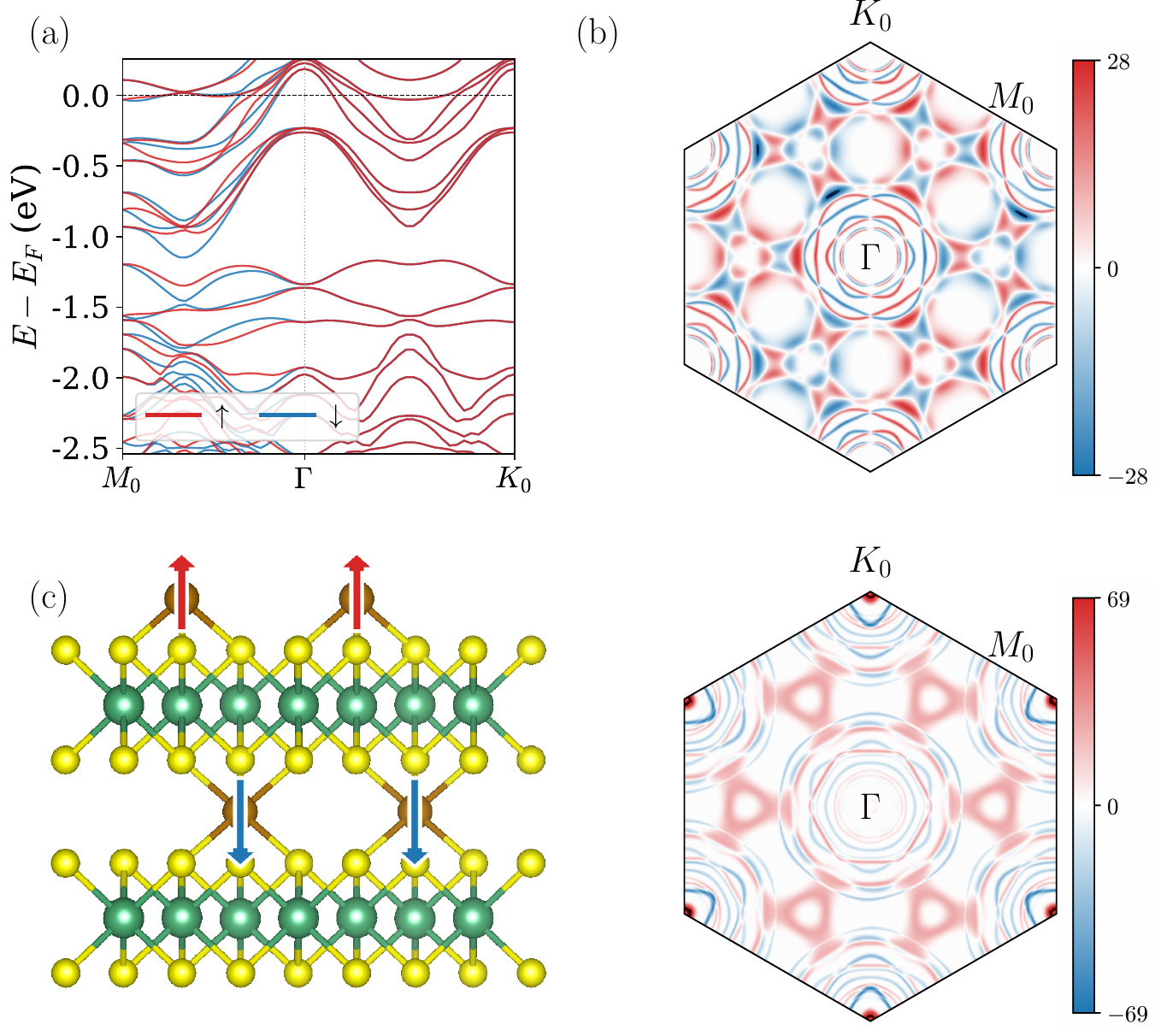}
    \caption{\justifying Spin splitting of A-type order in \(\mathrm{Fe}_{1/3}\mathrm{NbS}_{2}\). (a) Bulk A-AFM bands along \(M_0\)--\(\Gamma\)--\(K_0\) at \(k_z=\tfrac{5}{16}\tfrac{2\pi}{c}\) (spin projections along the moment axis: \(\uparrow\) red, \(\downarrow\) blue). (b) Fermi-level spin contrast \(\eta_z = A_\uparrow - A_\downarrow\) of the unfolded bulk spectral weight at the same \(k_z\). (c) Asymmetric Fe-terminated A-type slab and its Fermi-level spin contrast. This slab contains only the two-layer A-type unit, without the stripe-ordered layers used in Fig.~\ref{fig:2}. Maps in (b,c) are restricted to the primitive Brillouin zone.}
    \label{fig:altermag}
\end{figure}

Finally, we return to \(\mathrm{Fe}_{1/3}\mathrm{Nb}\mathrm{S}_{2}\) and
examine the symmetry consequence of its surface-stabilized A-AFM order. In bulk stripe order, opposite-spin
Fe sites are related by a lattice translation, so the bands are spin degenerate. In bulk A-type order, the opposite-spin layers are related only by the \(6_3\) screw rotation and twofold rotations, not by a pure translation or inversion, the defining symmetry of an altermagnet~\cite{Smejkal2022PRX040501,Smejkal2022PRX031042,day-roberts2026}. These rotations connect the spin-up and spin-down bands at different momenta, so their energies can differ at the same momentum. Without spin-orbit coupling, bulk A-type order therefore shows \(g\)-wave spin splitting [Figs.~\ref{fig:altermag}(a,b)], with bands split along
\(M_0\)--\(\Gamma\) and degenerate along \(\Gamma\)--\(K_0\). The splitting reaches \(55\)~meV
near the Fermi level at \(k_z=\tfrac{5}{16}\tfrac{2\pi}{c}\).

To examine the surface spin polarization, Fig.~\ref{fig:altermag}(c) uses an isolated Fe-terminated A-type bilayer as a simplified surface model, without the stripe-ordered interior layers. Its two opposite-spin layers are inequivalent and are not related by symmetry. This asymmetric bilayer exhibits nonrelativistic spin splitting and a Fermi-level spin contrast of both signs, together with a small net moment of (0.11~$\mu_B$) per ($\sqrt{3}\times\sqrt{3}$) cell. Furthermore, a symmetric A-type bilayer provides a symmetry control (Figs.~S2 and S3). There, an in-plane twofold rotation exchanges the layers and relates the spin-up and spin-down bands, producing sign-changing spin contrast with zero net moment. This control shows that spin splitting in an A-type bilayer does not require uncompensated magnetization. Together with the bulk symmetry analysis, these results suggest that surface-stabilized A-type order can exhibit altermagnetic character. Unlike surface altermagnetism produced by terminating an unchanged bulk order~\cite{lange2026}, here the surface changes the magnetic order itself. Spin-resolved ARPES~\cite{Krempasky2024Altermagnetism} can test the predicted surface spin polarization.

In summary, we have shown that the surface and bulk of one-third-intercalated TMDs can favor different magnetic orders. This extends earlier examples of surface-bulk magnetic differences in elemental and oxide magnets~\cite{Vescovo1993,Barbier2004,Hillebrecht2001} and goes beyond extraordinary surface phase transitions such as reported in CrSBr~\cite{Guo2024,Pei2024}. In
\(\mathrm{Fe}_{1/3}\mathrm{Nb}\mathrm{S}_{2}\), surface-sensitive ARPES is
reproduced by A-AFM order, whereas bulk neutron diffraction and energetics
give stripe order. Slab calculations reconcile the two by showing
that an Fe-terminated surface reverses the magnetic hierarchy and stabilizes A-AFM. The recovery of the \(K\)-centered Fermi-surface pockets in
\(\mathrm{Co}_{1/3}\mathrm{Nb}\mathrm{S}_{2}\) and
\(\mathrm{Co}_{1/3}\mathrm{Ta}\mathrm{S}_{2}\) suggests that related
surface-bulk dichotomy may extend to other magnetically intercalated TMDs. In
\(\mathrm{Fe}_{1/3}\mathrm{Nb}\mathrm{S}_{2}\), the surface-stabilized
A-AFM phase carries nonrelativistic, sign-changing
spin splitting, linking the magnetic dichotomy to altermagnetism.
More broadly, surface-bulk magnetic dichotomy provides a route to realizing near-surface magnetic phases and electronic functionality distinct from the bulk ground state in van der Waals magnets. 

\nocite{ku2010,popescu2012}

\begin{acknowledgments}
\emph{Acknowledgments} -- K.L. and T.Z. were supported by the National Science Foundation under Grant No.~CHE-2337991. W.L., K.D., and Y.H. were supported by the U.S. Air Force Office of Scientific Research under Award No.~FA9550-24-1-0048. The authors acknowledge additional support from the Yale AI seed grant. The work at SCU was supported by the College of Arts and Sciences (CAS) Grant at Santa Clara University. S.W. also acknowledges support from the Clare Boothe Luce Scholar STEM Grant. This work used Bridges-2 at the Pittsburgh Supercomputing Center through allocation CHE260011 from the Advanced Cyberinfrastructure Coordination Ecosystem: Services \& Support (ACCESS) program, which is supported by U.S. National Science Foundation grants \#2138259, \#2138286, \#2138307, \#2137603, and \#2138296. The authors thank the Yale Center for Research Computing for providing additional computing resources. Bulk crystal growth of NbS$_2$ materials was supported by the 2DCC-MIP under NSF cooperative agreement DMR-2039351.
We used OpenAI GPT (5.6–6.1 Sol and 6 Astra) models to set up and post-process calculations and edit the manuscript. The authors directed all AI-assisted work, verified its outputs, and take full responsibility for the content.
\end{acknowledgments}

\bibliography{ai4arpes,add_ref}

\end{document}